# Thermal Conductivity of Isotopically Enriched $^{28}$Si: Revisited

R. K. Kremer, K. Graf, M. Cardona
*Max-Planck-Institut für Festkörperforschung, D-70569 Stuttgart, Germany*
G.G. Devyatykh, A.V. Gusev, A.M. Gibin
Institute of Chemistry of High-Purity Substances, Russian Academy of Sciences, 603950, Nizhny Novgorod, Russia.
A. V. Inyushkin, A. N. Taldenkov
*Russian Research Centre Kurchatov Institute, 123182 Moscow, Russia*
H.-J. Pohl
*VITCON Projectconsult GmbH, 07745 Jena, Germany*

The thermal conductivity of isotopically enriched $^{28}$Si (enrichment better than 99.9%) was redetermined independently in three laboratories by high precision experiments on a total of 4 samples of different shape and degree of isotope enrichment in the range from 5 to 300 K with particular emphasis on the range near room temperature. The results obtained in the different laboratories are in good agreement with each other. They indicate that at room temperature the thermal conductivity of isotopically enriched $^{28}$Si exceeds the thermal conductivity of Si with a natural, unmodified isotope mixture by 10±2 %. This finding is in disagreement with an earlier report by Ruf *et al.* At ~26 K the thermal conductivity of $^{28}$Si reaches a maximum. The maximum value depends on sample shape and the degree of isotope enrichment and exceeds the thermal conductivity of natural Si by a factor of ~8 for a 99.982% $^{28}$Si enriched sample. The thermal conductivity of Si with natural isotope composition is consistently found to be ~3% lower than the values recommended in the literature.

## INTRODUCTION

Phonon scattering due to the presence of different isotopes in an otherwise pure crystal (no chemical defects or dislocations) has been identified as a mechanism that strongly affects the thermal conductivity $\kappa$ [1–3]. The availability of larger quantities of highly isotope-enriched materials recently revived the interest in this effect in order to study the mechanisms underlying the thermal conductivity and possibly to improve material properties [4–8].

Recently, the observation of a significantly enhanced thermal conductivity of isotopically enriched $^{28}$Si near room temperature by Capinski [9] and Ruf et al. [10] has generated large interest. In these studies, it was found that the thermal conductivity of isotopically enriched $^{28}$Si is enhanced by about 60% over that of silicon with natural isotopic composition. This rather high isotope effect attracted considerable attention concerning both, fundamental physics and applications. For technical applications, a significantly enlarged thermal conductivity at room temperature would be of interest for high performance electronic devices. [11]. From the fundamental physics aspect, the experimental results were unexpected since the observed isotopic effect was significantly larger than the prediction of simple theoretical estimates [10–13] and more advanced model calculations [14]. On the other hand, theoretical papers [15–17] were published with the results supporting the data of refs. [9,10]. A large isotopic effect at room temperature is in principle only possible if normal phonon-phonon scattering processes play an important role in determining the formation of the non-equilibrium distribution function of phonons at these temperatures.

The large room temperature isotope effect was questioned following the results of an experimental study by Gusev *et al.* [18] which indicate that the room temperature isotopic effect amounts to only 7%, and more recently by an erratum in which an increase at room temperature of about 10% was reported [19]. Thus, presently there is considerable uncertainty concerning the effect of isotope disorder on the thermal conductivity of silicon, especially in the technologically relevant regime around room temperature. This and the broad interest for potential technical applications led us to carefully redetermine the thermal conductivity of $^{28}$Si in comparison with that of Si with the natural isotope composition ($^{nat}$Si). For this purpose, absolute measurements of the temperature dependence of the thermal conductivity of several samples of $^{28}$Si and of natural silicon using different experimental techniques, were carried out independently in three laboratories, in Stuttgart, Nizhny Novgorod and Moscow. Here we report the results of these experiments. All experiments conclusively find that at room temperature the thermal conductivity of $^{28}$Si exceeds that of $^{nat}$Si by only 10% while at the maximum close to 26 K, $\kappa$ is enhanced by almost an order of magnitude.

## EXPERIMENTAL DETAILS

Thermal conductivities were measured on a total of four different bar-shaped samples of enriched $^{28}$Si, (99.9% and higher) with a broad range of cross sections ($\sim 4 - 20$ mm$^2$) and lengths (20 – 50 mm). The samples were cut from crystals grown by the floating-zone method such that the orientation along the sample length was always [100]. A detailed description of the preparation of isotopically enriched $^{28}$Si crystals has been given elsewhere [20]. The samples with natural isotopic composition, which were measured for comparison, had the same sizes and crystallographic orientation and similar chemical purity. Chemical, isotope and geometrical details of all investigated samples of $^{28}$Si are compiled in Table I. The surfaces of samples $^{28}$Si-M and $^{28}$Si-NN were particularly ground with 14 μm abrasive powder slurry to ensure diffuse scattering of thermal phonons from the sample surfaces at low temperatures.

In all three laboratories, the thermal conductivities were measured by the steady-state heat flow technique using either platinum resistance thermometers (Lake Shore Cryotronics, Inc.) (50–320 K) or Cernox thermometers (CX−1050 Lake Shore) (5−300 K) as temperature sensors and a surface mounted thick film (SMD) resistor as a heater which was glued to one end face of the samples. The opposite end of the sample bars was thermally anchored to a Cu block which was thermally coupled to the cryostat. Distances between the temperature sensors were $\Delta x \approx 10$–20 mm. After sufficient stabilization to a particular temperature $T$, the electrical power $P$ dissipated in the SMD heater was adjusted such that typical temperature gradients $\Delta T$ amounted to 0.01–0.2 K (Kurchatov Institute and IChHPS RAS) while in Stuttgart several (typically 4–6) stable gradients $0.01 \leq \Delta T \leq 0.7$ K were generated and the thermal conductance was obtained from a least-squares fit of the set of measured ($P_i$, $\Delta T_i$) pairs. The latter approach provides additional signal-to-noise improvement and enables a critical assessment of deviations from the linear relationship $\Delta T_i \propto P_i$ and thus allows us to recognize self-heating of the sample due to possible insufficient thermal coupling to the bath. The thermal conductivity $\kappa(T)$ at an average temperature $(T_1+T_2)/2$ in turn is calculated with the equation $P = \kappa(T)\, A\, \Delta T/\Delta x$, where $\Delta T = (T_1 - T_2)$ is the stable temperature gradient measured between the two thermometers at a distance $\Delta x$ apart in the presence of the constant heat flow $P=U \cdot I$, $I$ being the current and $U$ the voltage drop across the SMD resistance heater. $A$ is the cross section of the sample. The samples were mounted into evacuated Cu cans which were immersed into either liquid nitrogen or liquid He (IChHPS RAS, Kurchatov Institute) or mounted into a continuous flow He evaporation cryostat (MPI FKF).

The temperature sensors were either attached to the sample bars by Cu clamps with an indium layer to improve the thermal contact (Kurchatov Institute and IChHPS RAS) or by glueing either the Cernox or Pt chip thermometers directly to the sample using an epoxy resin (MPI FKF). The absolute errors in the determination of the thermal conductivity arise from the finite thickness of the temperature sensors or the clamps employed to connect the sensors to the samples. For both methods we estimate the absolute errors to 2–3%. Particular emphasis was given to using the same thermometer mounting when measuring the isotopically enriched and the natural Si samples. This enables us to reduce the relative error by a factor of ~2 when comparing both types of samples. Towards higher temperatures, thermal radiation opens an additional path to dissipate the supplied heat. Thermal radiation increases with temperature like $T^3 \Delta T$ and, compared to thermal conduction, it becomes increasingly important for samples with smaller cross sections and comparatively larger surface areas. For the sample with the smallest cross section investigated (MPI FKF, $^{28}$Si-S1, $A$~3.8 mm$^2$ ) assuming an infrared emissivity of ~0.5, radiation losses were estimated to contribute by about 2% to the heat dissipation at room temperature. Thermal radiation losses at room temperature are therefore smaller than other experimental errors and become negligible for the samples with larger cross sections. To test the resolution and eventually correct for radiation losses, the thermal conductivity of isotopically enriched samples with different cross sections was determined with especially high precision around room temperature and above.

**RESULTS AND DISCUSSION**

The thermal conductivities $\kappa(T)$ of three isotopically enriched $^{28}$Si samples as measured independently by the three laboratories are displayed in Fig. 1 in comparison to the thermal conductivities of Si with the natural isotope composition and to the literature data [21]. For all sets of data, the thermal conductivity of $^{28}$Si at 300 K is 10±2% larger than that of $^{nat}$Si. This finding is in very good agreement with the results of Morelli *et al.* who predicted an increase at room temperature of 12% [14]. Above 80 K, the results of the different laboratories agree to within 3%. At lower temperatures, we observe characteristic differences, which can be ascribed to the different degrees of isotope enrichment, chemical purity and, at the lowest temperatures, to the different sample dimensions and surface finish. Comparison of the thermal conductivity of $^{nat}$Si with the standard data recommended by Touloukian [21] additionally reveals that at room temperature our three laboratories consistently find the thermal conductivity of $^{nat}$Si to be about 3% smaller than the standard values. Figure 2 displays the inverse thermal conductivity (‘thermal resistance’) of $^{nat}$Si together with a polynomial of order two as guide for the eyes. In the temperature range 220 K $\leq T \leq$320 K the thermal resistance, within error bars, closely follows the equation $\kappa^{-1}$[m K W$^{-1}$] = $1.60(2)\times10^{-5}\ T + 2.37(6)\times10^{-8}\ T^2$ with coefficients similar to those derived by Glasbrenner and Slack [17], however, truncating a temperature independent contribution of ~$3\times10^{-4}$ mK W$^{-1}$ as used in [13].

The low-temperature data displayed in Fig. 3 reflect the degree of isotope enrichment, chemical purity and surface conditions of the different samples under investigation. At low temperatures, the thermal conductivity approaches a $T^3$ power law behavior, however, with slightly different prefactors for the different samples. While at these temperatures the thermal conductivity of the $^{28}$Si sample investigated at the Kurchatov Institute and at the IChHPS RAS coincides with that of $^{nat}$Si, the thermal conductivities of the MPI FKF $^{28}$Si sample are by about a factor of 2 larger than those of $^{nat}Si$. We ascribe this difference to additional phonon scattering by chemical impurities or different surface scattering due to different sample surface finishing [7]. We finally discuss the thermal radiation losses which around room temperature may provide an additional channel for heat dissipation in samples with a small cross section. Figure 4 displays the "nominal" thermal conductivities of two samples which differ in cross section by about a factor of 5. The total sample length and the distance between the temperature sensors were approximately the same. We can clearly observe an increasing difference of the "nominal" thermal conductivity between the two samples which grows approximately $\propto T^3$ thus indicating that the difference is due to increased radiation losses for the sample with smaller cross section. Using the Stefan-Boltzmann law for thermal radiation, we can estimate the ratio of heat flow due to thermal radiation $P_{\mathrm{rad}}$ and conduction $P_{\mathrm{cond}}$ to be $P_{rad} / P_{cond} = \sigma\ \varepsilon\ T^3\ l\, S /( A\, \kappa(T))$, where $S$ is the sample surface, $A$ the cross section, $l$ the sample length, $\varepsilon$ the infrared emissivity and $\sigma$ the Stefan-Boltzmann radiation constant. Using the sample dimensions and assuming an average infrared emissivity $\varepsilon$~0.5, we estimate that for the sample with cross section $A$~4 mm$^2$ the radiation losses at room temperature amount to about 2%, which is in very good agreement with the experimental observation. Consequently, for the larger sample of ($A$~20 mm$^2$) radiation losses at room temperature are negligible.

In summary, in a collaboration involving three laboratories, we have independently redetermined the thermal conductivity of natural Si and isotopically enriched $^{28}Si$ samples. In the temperature range 80 K<$T$<300 K, we find excellent agreement between the three sets of results. We therefore conclude that at room temperature the thermal conductivity $\kappa$ of $^{28}Si$ exceeds $\kappa$ of $^{nat}Si$ by only 10±2 %. Optimized experimental conditions, improved samples and the close coincidence of all experimental results indicate that the original measurements by Ruf *et al.* represent an overestimate of $\kappa(T)$ at room temperature, a fact which has already been admitted by these authors [19]. In addition, we find that close to room temperature the values of the thermal conductivity of $^{nat}Si$ are about 3% lower than those recommended in the existing literature.

*Acknowledgement*

We acknowledge S. Höhn for valuable experimental assistance.

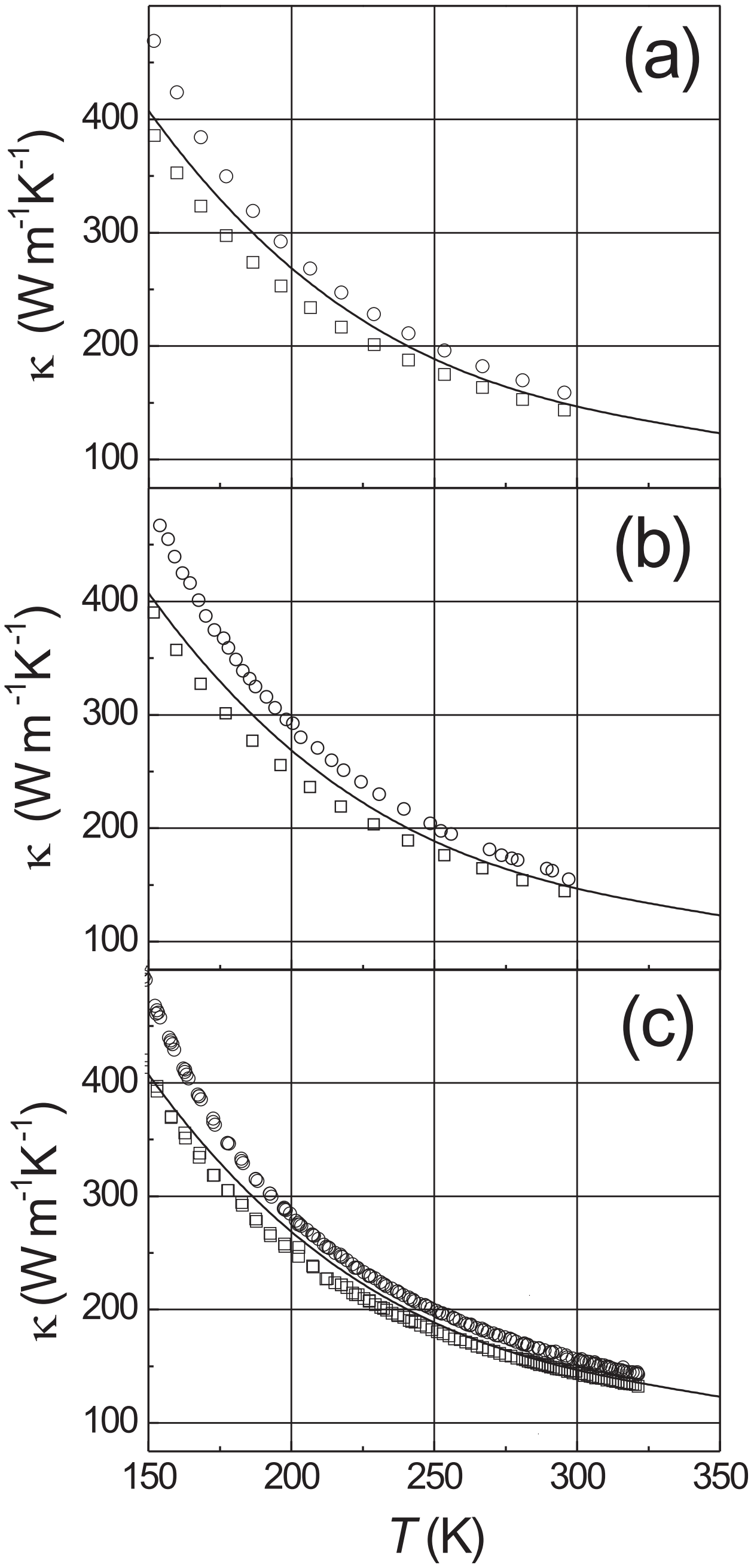

Fig. 1 Thermal conductivities (○) of three samples of $^{28}$Si. (a) sample $^{28}$Si-M, (b) sample $^{28}$Si-NN and (c) sample $^{28}$Si-S2 (sample details cf. Table 1) in comparison with the thermal conductivities of samples of $^{nat}$Si with similar sample sizes (~). The solid lines represent the thermal conductivity of $^{nat}$Si given in [21].

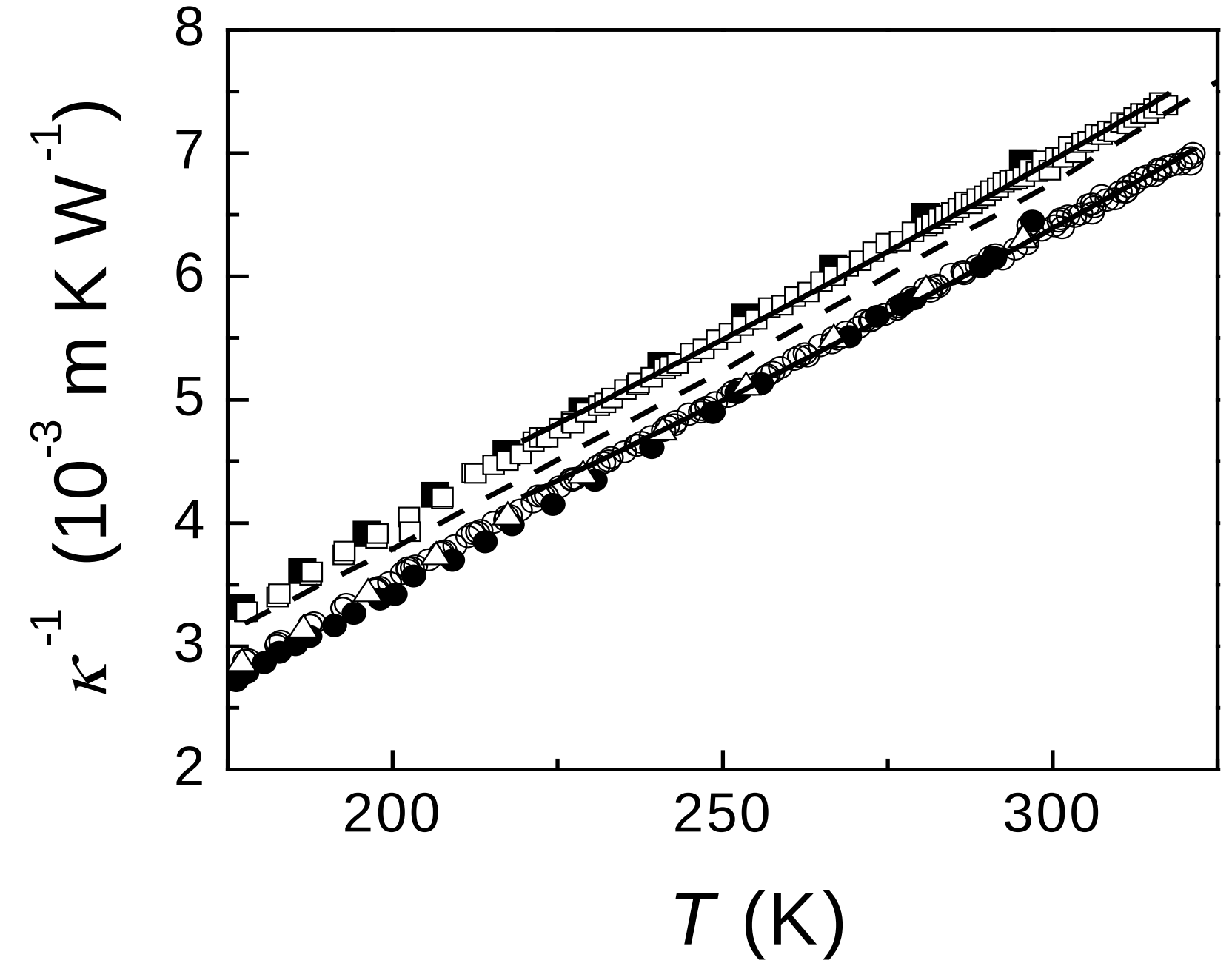

Fig. 2 Thermal resistivities of three samples of $^{28}$Si ( ○ $^{28}$Si-S2, ● $^{28}$Si-NN, ρ $^{28}$Si-M) in comparison with the thermal resistivity of $^{nat}$Si (■,□). The full lines are fits with 2$^{nd}$ order polynomials in the temperature range 220 K<*T*<320 K with parameters: $^{28}$Si: $\kappa^{-1}$ [m W K$^{-1}$]= 1.33(2)×10$^{-5}$ $T$ + 2.66(5)×10$^{-8}$ $T^{2}$; $^{nat}$Si: $\kappa^{-1}$ [mW K$^{-1}$]=1.60(2)×10$^{-5}$ $T$ + 2.37(6)×10$^{-8}$ $T^{2}$. The dashed line represents the standard data for $^{nat}$Si compiled in [21].

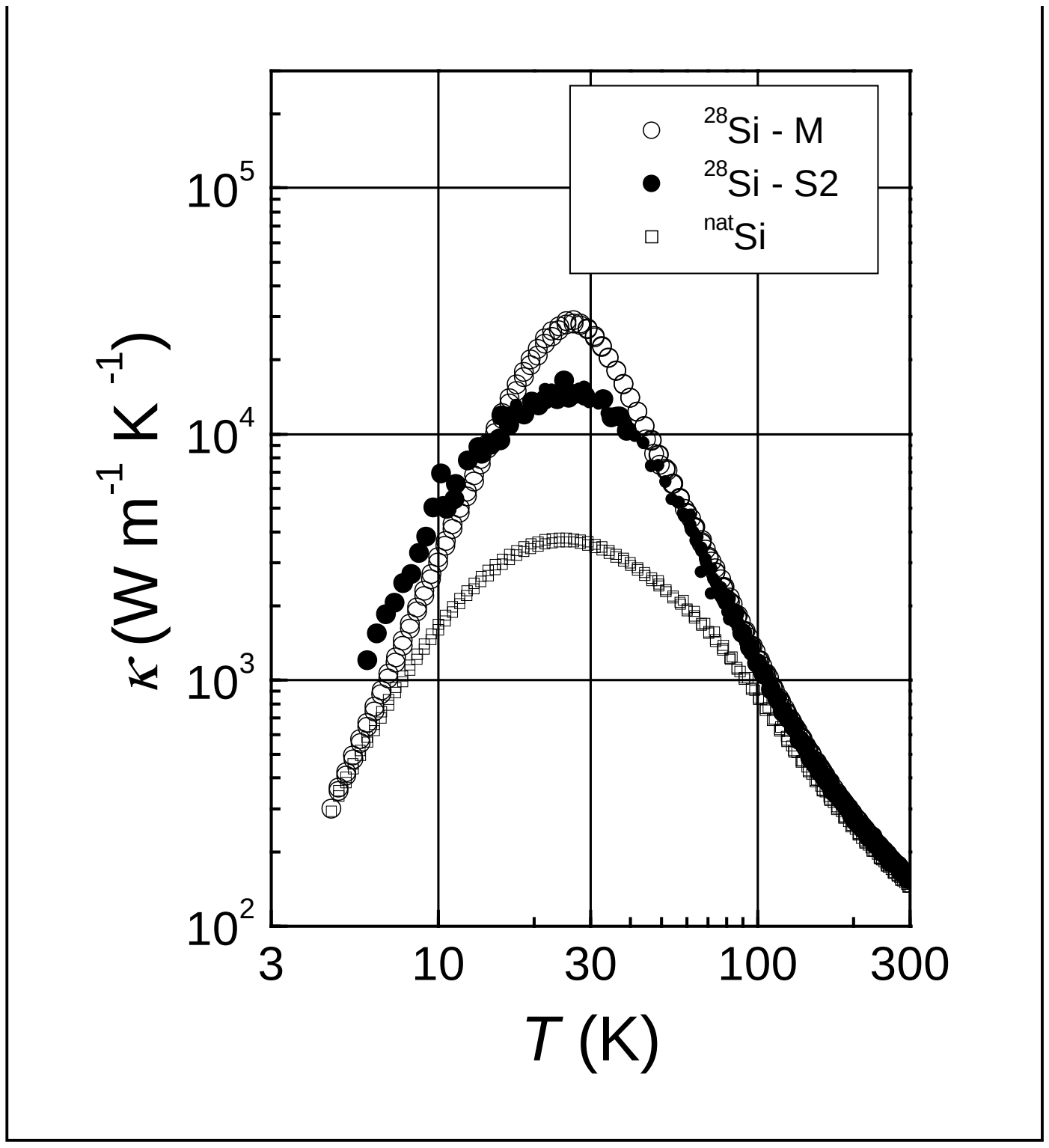

Fig. 3 Thermal conductivity of two samples of $^{28}$Si (○ $^{28}$Si-M, ● $^{28}$Si-S2) in comparison with the thermal conductivity of $^{nat}$Si with size and surface finish similar to sample $^{28}$Si-M.

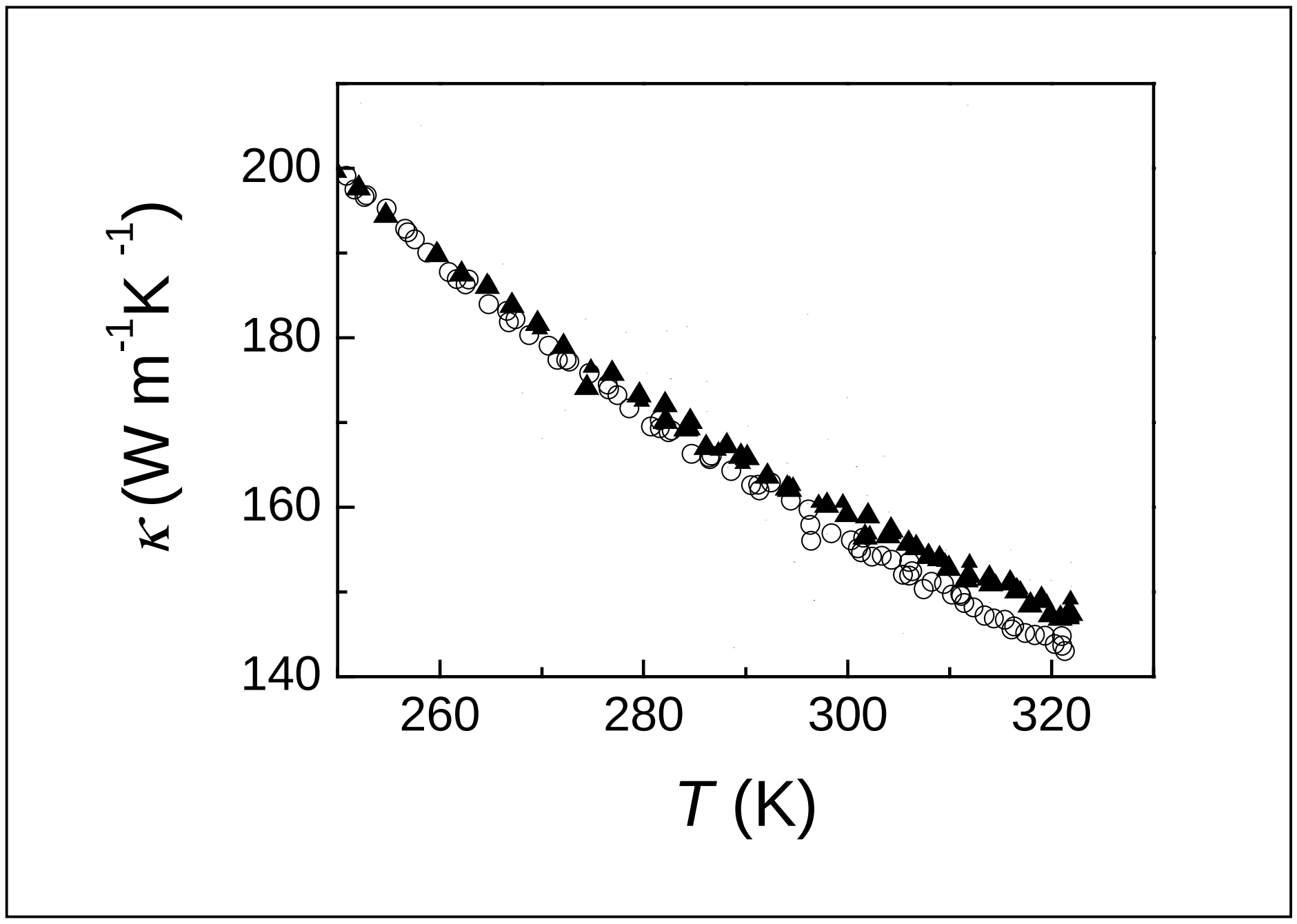

Fig. 4 Thermal conductivity of two samples of $^{28}$Si with cross section 3.8 mm$^2$ (▲, sample $^{28}$Si-S1) and 19.6 mm$^2$ (○, sample $^{28}$Si-S2).

| sample | isotopic composition (%) | | | orien-tation | content of impurities | | temperature range ( K) | size ($mm^3$ ) |
|---|---|---|---|---|---|---|---|---|
| | $^{28}Si$ | $^{29}Si$ | $^{30}Si$ | | carbon, oxygen ($cm^{-3}$) | metals ($cm^{-3}$) | | |
| $^{28}$Si-M | 99.9829 | 0.0144 | 0.0027 | [100] | $2\times10^{15}$ – C<br>$1\times10^{15}$ – O | $<10^{-5}$ | 5 - 300 | $2.5 \times 2.0 \times 23$ |
| $^{28}$Si-S1 | 99.859 | ~0.13 | ~0.02 | [100] | $5\times10^{16}$ – C<br>$2\times10^{16}$ – O | $<10^{-5}$ | 50 - 320 | $2.0 \times 1.9 \times 60$ |
| $^{28}$Si-S2 | 99.979 | ~0.019 | ~0.002 | [100] | $5\times10^{16}$ – C<br>$2\times10^{17}$ – O | $<10^{-5}$ | 6 – 320 | $4.5 \times 4.35 \times 60$ |
| $^{28}$Si-NN | 99.9829 | 0.0144 | 0.0027 | [100] | $2\times10^{15}$ – C<br>$1\times10^{15}$ – O | $<10^{-5}$ | 5 - 300 | $3.12 \times 2.0 \times 20$ |

Table 1 Physical and chemical parameters of the $^{28}$Si samples measured at RRC, Kurchatov Institute (Moscow) ( $^{28}$Si-M), at the MPI FKF (Stuttgart)( $^{28}$Si-S1, $^{28}$Si-S2) and at the IChHPS RAS (Nizhny Novgorod) ( $^{28}$Si-NN).